\documentstyle[12pt,graphicx]{article}
\begin{document}
\title{Vacuum Polarization of a Charged Massless Scalar
Field on Cosmic String Spacetime \\
in the Presence of a Magnetic Field}
\author{J. Spinelly {\thanks{E-mail: spinelly@fisica.ufpb.br}}  
and E. R. Bezerra de Mello \thanks{E-mail: emello@fisica.ufpb.br}\\
Departamento de F\'{\i}sica-CCEN\\
Universidade Federal da Para\'{\i}ba\\
58.059-970, J. Pessoa, PB\\
C. Postal 5.008\\
Brazil}
\maketitle
\begin {abstract}

In this paper we consider a charged massless scalar quantum field operator
in the spacetime of an idealized cosmic string, i.e., an infinitely long, straight 
and static cosmic string, which presents a magnetic field confined in a 
cylindrical tube of finite radius. Three distinct situations are taking 
into account in this analysis: {\it{i)}} a
homogeneous field inside the tube, {\it{ii)}} a magnetic field proportional to 
$1/r$  and {\it{iii)}} a cylindrical shell with $\delta$-function. In these 
three cases the axis of the infinitely long tube of radius $R$ coincides with 
the cosmic string. In order to study the vacuum polarization effects outside 
the tube, we explicitly calculate the Euclidean Green function associated with 
this system for the three above situations, 
considering points in the region outside the tube. Having these Green 
functions we calculate the renormalized vacuum expectation values, 
$\langle\hat{\Phi}^*(x)\hat{\Phi}(x)\rangle_{Ren.}$ and 
$\langle\hat{T}_\mu^\nu(x)\rangle_{Ren.}$, associated with the charged field. In 
the evaluation of these vacuum polarization effects, two contributions appear for 
the three models. The firsts are the standard ones due to the conical 
geometry of the spacetime and the magnetic flux. The seconds contributions appear as an 
extra term. They are corrections due to the finite thickness
of the radius of the tube. These extra terms provide relevant contributions, 
even for points very far way from the system, like a long-range effect. 
\end{abstract}

\newpage
\renewcommand{\thesection}{\arabic{section}.}

\section{Introduction}
One of the firsts models which present topological defects was investigated
by Nielsen and Olesen \cite{N-O} several years ago. In their analysis
topologically stable solutions named vortices, appears as a consequence
of spontaneous breakdown of local Abelian and non-Abelian gauge symmetries
in a classical field theory. These solutions correspond to infinitely 
long objects which carry magnetic fluxes. Later, Garfinkle investigated the 
Abelian model considering its influence on the geometry of the spacetime 
\cite{Garfinkle}. He found, as in flat spacetime, static cylindrically 
symmetric solutions representing vortices. He also shown that asymptotically 
the spacetime around the cosmic string is a Minkowiski one minus a wedge. Its 
core has a non-zero thickness, and the magnetic field vanishes outside of it. 
Two years later Linet \cite{Linet} obtained, under some specific condition, 
exact solution for the metric tensor. He was able to show that the structure 
of the respective spacetime corresponds to a conical one, with the conicity 
parameter being expressed in terms of the energy 
per unity length of the vortex. The complete analysis about the behavior of 
the gauge and matter fields near the cosmic string's core can only be obtained 
numerically. Some recent numerical analysis \cite{Yves} about the structure of 
supermassive cosmic strings show that two different kind of solutions for the 
metric tensor exist. 

A complete analysis about the behavior of a charged field in the neighborhood
of a $U(1)-$gauge cosmic string, must take into account not only the influence 
of the geometry of the spacetime but, also, the influence of the magnetic
field. Two distinct cases can be analysed: $i)$ The first one is to consider
the string as an idealized infinitely thin linear topological defect, having a 
magnetic field running along it. This case can be treated analytically.
$ii)$ The second approach is to consider the non-zero thickness for the
string. Analytically this problem becomes intractable. An intermediate
approach is what we shall adopt here. We shall use an approximated model 
considering the spacetime produced by the string as being conical
everywhere, but having a magnetic field inside a tube of non-zero thickness 
surrounding it. In this way some improvement is introduced when compared with 
the ideal case. Moreover, we shall admit specifics spatial behaviors for the 
magnetic field inside the tube trying to approach the real situation.

The vacuum polarization effects due to a magnetic field confined in a tube of
finite radius in Minkowski spacetime has been first
analysed by Serebryanyi \cite{S}. A few years later this effect in a charged 
massless scalar field on a idealized cosmic string spacetime, has been 
calculated by Guimar\~aes and Linet \cite{linet1}. There a magnetic flux 
running through the line singularity was considered. As a consequence the 
renormalized vacuum expectation value 
associated with the energy-momentum tensor, $\langle\hat{T}_\mu^\nu(x)
\rangle_{Ren.}$, presents a contribution coming from the geometry of the
spacetime and the magnetic flux. In order to obtain this result the respective 
Green function was determined. In a particular choice of gauge, this Green 
function depends explicitly on the factor $\gamma$, the fractional part of the 
ratio of the magnetic flux $\phi$ by the quantum flux, $\phi_0=2\pi/e$.

More recently Sriramkumar \cite{Sri} has calculated the vacuum fluctuation of 
current and energy densities for a massless charged scalar field around an 
idealized cosmic string carrying a magnetic flux. His approach to obtain the 
Green function takes into account the presence of the vector potential in the
differential operator, $D_\mu=\partial_\mu-ieA_\mu$, which presents the 
advantage to calculate the two-points function without imposing any additional 
boundary condition on the field. Also the vacuum polarization associated
with massless scalar field in a infinitely thin cosmic string embedded in a 
Randall-Sundrum type $2$ brane world, has been calculated \cite{Grats}. In
lowest order of perturbation, $\langle T_{\mu\nu}\rangle_{Ren.}$, consists of
two parts. The first one depends only on the conicity parameter and the 
second is a correction which depends on the parameter $k$ connected with the
cosmological constant.

Allen, Kay and Ottewill \cite{allen} have analyzed the vacuum 
polarization effect of a massless scalar field on a realistic cosmic string 
spacetime, considering generically the effect of the string's core through 
the non-minimal coupling between the scalar field with the geometry: 
$\xi {\cal{R}}$, being ${\cal{R}}$ the Ricci scalar and $\xi$ the coupling
constant \footnote{In \cite{allen} was not considered a specific model for the 
inner structure of the string; however, it was considered be infinitely long, 
straight and static.}. In this case the respective Green function presents 
two parts: The first one, named "regular" by the authors, coincides with the usual 
one obtained in the literature for an idealized cosmic string, see \cite{smith} 
for example. The second contribution comes from the nonzero core's radius of the 
string. Although the radius of the cosmic string's core is very small
\footnote{For cosmic string formed by a typical grand unified theory is of order 
$10^{-30}$ cm.}, the effect of this finite thickness on the vacuum polarization 
of a massless scalar field, is a long-range effects. Moreover it is of the same 
order of magnitude as contribution coming from the geometry up to distance from 
the cosmic string that exceed the radius of the observable Universe.

Motivated by this result we raised a question about the consequence on the vacuum 
polarization of a charged massless scalar filed in the $U(1)-$gauge cosmic string 
spacetime considering this string surrounded by a magnetic filed confined in a 
tube of finite radius. As in the Allen {\it et al} paper, we expect some 
consequence of the finite thickness of the tube on the vacuum polarization
effect. In order to investigate this subject we decided in this paper, for
simplicity, to analyse the behavior of a quantum charged massless scalar field 
in a idealized cosmic string spacetime, considering the presence of a infinitely
long tube of finite radius $R$ of magnetic field. Three distinct cases were 
taking into account:\\ 
{\it{i)}} a homogeneous magnetic field inside the tube, \\
{\it{ii)}} a magnetic field proportional to $1/r$ inside and \\
{\it{iii)}} a magnetic field on a cylindrical 
shell. \\
In all these three cases the axis of the infinitely long and straight tube 
coincides with the cosmic string.

The idealized model for a infinitely long straight static cosmic string spacetime 
can be given in cylindrical coordinates by the line element below:
\begin{equation}
ds^2=-dt^2+dr^2+\alpha^2 r^2d\theta^2+dz^2 \ , 
\label{1}
\end{equation}
where $\alpha$ is a parameter smaller than unity which codify the presence of a 
conical two-surface $\left( r,\theta \right)$. In fact for a typical Grand Unified 
Theory, $\alpha=1-O(10^{-6})$. (Here we considered that this cosmic string is 
lying along the $z$-direction.)

A magnetic field along the $z$-direction can be represented by ${\vec{H}}(r)=H(r)
\hat{z}$. Now assuming that the field has a finite range in the radial coordinate, 
we are particularly interested in the three models below:
\begin{eqnarray}
\label{2}
i)\ \  H(r)&=&\frac{\phi}{\alpha\pi R^2}\Theta(R-r),\ \ \mbox{homogeneous field 
inside} 
\\
\label{3} 
ii)\ \  H(r)&=&\frac{\phi}{2\pi\alpha Rr}\Theta(R-r),\ \ \mbox{field proportional 
to 
$1/r$ inside} 
\\
\label{4}
iii)\ \  H(r)&=&\frac{\phi}{2\pi\alpha R}\delta(r-R),\ \ 
\mbox{cylindrical shell}, 
\end{eqnarray}
where $R$ is the radial extent of the tube, and $\phi$ is the total flux. The 
ratio of the flux by the quantum flux $\phi_{o}$, can be expressed by 
$\phi/\phi_0=N+\gamma$, where $N$ is the integer part and $0<\gamma<1.$
\footnote{These three models for the magnetic field has been considered
 previously by Bordag and Voropaev \cite{bordag} to analyze the quantum 
mechanical behavior of spin$-1/2$ and gyromagnetic $g\neq 2$ particle in a 
field of a magnetic string.} 

Considering both effects, due to the geometry of the spacetime and the 
finite thickness of the tube of magnetic field on 
the vacuum polarization of a charged massless scalar field, we observe that 
there appear two distinct contributions for each case: For the cases $({\it{i}})$ and $({\it{ii}})$, the firsts ones are due to 
the geometry and the magnetic flux. They are similar to the result found by 
Guimar\~aes and Linet \cite{linet1}. The seconds contributions are due the 
nonzero radius of the tube, $R$. The latter present similar
dependence with the radial coordinate. In fact their dependence on the ratio
$R/r$ is of power type besides an overall $1/r^4$ dependence. These results
indicate that the detailed information 
about the structure of the magnetic field inside the tube is not so relevant. The 
third case is completely different from the others. The second contribution to 
the vacuum polarization is now very similar to the result obtained in 
\cite{allen}. It presents a logarithmic dependence on the radial 
coordinate, being a long-range contribution.

Taking into account that the topological defects are of evident physical 
interest (they are considered as good candidates to explain some components
of anisotropy on the cosmic microwave background \cite{Sarangi}, gamma ray
burst \cite{Berezinski}, gravitational waves \cite{Damour} and highest
energy cosmic ray \cite{Bhattacharjee}) it seems reasonable to investigate
the above mentioned problem in more detail.

This paper is organized as follows. In the section $2$, we explicitly calculate 
the Green functions associated with this system for the three different models.
Having now these Green functions, in section $3$ we explicitly
calculate the renormalized vacuum expectation values, 
$\langle\hat{\Phi}^*(x)\hat{\Phi}(x)\rangle_{Ren.}$, in complete form. Using a 
judicious criteria of approximation we present the behavior for these 
quantities. In section $4$, we calculate the renormalized vacuum expectation values
of the energy-momentum operator, $\langle\hat{T}_\mu^\nu(x)\rangle_{Ren.}$. 
Using the same procedure as in the previous section we present a approximated expressions for these results. We leave for section $5$ our conclusions and 
most relevant remarks about this paper. Finally we want to say that we shall
use in this paper $\hbar=G=c=1.$

\section{Green Function}

The Green function associated with a charged massless scalar field 
must obey the non-homogeneous second order differential equation below
\begin{equation}
\label{Gr}
\frac{1}{\sqrt{-g}}{\mathcal{D}}_{\mu}\left[\sqrt{-g}g^{\mu\nu}{\mathcal{D}}_{\nu}
\right]G(x,x^{'})=-\delta^{(4)}(x-x'),
\label{5}
\end{equation} 
where ${\mathcal{D}}_{\mu}=\partial_\mu-ieA_\mu$.

Because it is easier to calculate the vacuum average of the square of the
field and the energy-momentum tensor operators using the Euclidean version
of the Green function, let us solve the above equation adopting the 
Euclidean extension of the cosmic string metric tensor given in (\ref{1}), by
performing a Wick rotation $t \rightarrow i\tau$. Now if we want to represent a
magnetic field along the $z$-direction, the four-vector potential reads
\begin{equation}
A_{\mu}=(0,0,A(r),0) \ .
\label{6}
\end{equation} 
So the equation (\ref{5}) becomes 
\begin{eqnarray}
\left[\frac{\partial^{2}}{\partial\tau^2}+\frac{1}{r}\frac{\partial}{\partial r}
\left(r\frac{\partial}{\partial r}\right)+\frac{\partial^2}{\partial z^2}+
\frac{1}{\alpha^2 r^2}\left(\frac{\partial^2}{\partial \theta^2}-2ieA
\frac{\partial}{\partial \theta}-e^2A^2 \right) \right]G(x,x')=
\nonumber\\
-\frac{1}{\alpha r}\delta(\tau-\tau')\delta(r-r')\delta(\theta-\theta')
\delta(z-z') \ .
\label{7}
\end{eqnarray}

Now, in order to reproduce the magnetic fields configurations given by 
(\ref{2})-(\ref{4}), we have only to write the third component of the vector
potential by
\begin{equation}
A=\frac{\phi}{2\pi\alpha}a(r) \ .
\label{8} 
\end{equation}

For the two firsts models, we can represent the radial function $a(r)$ by:
\begin{equation}
a(r)=f(r)\Theta (R-r)+\Theta (r-R) \ ,
\label{9}
\end{equation}
with
\begin{eqnarray}
f(r)=\left\{\begin{array}{cc}
r^2/R^2,&\mbox{for the model ({\it{i}}) and}\\
r/R,&\mbox{for the model ({\it{ii}}).}
\end{array}
\right.
\label{10}
\end{eqnarray}
For the third model,
\begin{equation}
a(r)=\Theta(R-r).
\label{10.1}
\end{equation}
Due to the cylindrical symmetry of this system the Euclidean Green function can 
be expressed by
\begin{equation}
G(x,x')=\frac1{(2\pi)^3}\sum_{n=-\infty}^\infty e^{in(\theta-\theta')}
\int_{-\infty}^\infty dk e^{ik(z-z')} \int_{-\infty}^{\infty} d\omega 
e^{i\omega (\tau-\tau')} g_n(r,r') \ .
\label{11}
\end{equation}

Before to specialize on the specific model let us write down the non-homogeneous
differential equation obeyed by the unknown function $g_n(r,r')$.
Substituting (\ref{11}) into (\ref{7}) and using the standard representation to 
the delta functions in the temporal, angular and $z$-coordinates, we arrive to 
following differential equation for the function $g_{n}(r,r')$:
\begin{equation}
\left[\frac{d^2}{dr^2}+\frac{1}{r}\frac d{dr}-\frac1{\alpha^2r^2}(n^2-2neA+
e^2A^2)-\beta^2 \right]g_{n}(r,r')=-\frac1{\alpha r}\delta(r-r') \ ,
\label{12}
\end{equation}
where $\beta^2=k^2+\omega^2$.

So, all the informations about the magnetic field, the radius of the cylindrical
tube and the conicity of the spacetime are contained in $g_n(r,r')$. However,
before to obtain specific solution, corresponding to specific model of
magnetic field, some general properties must be satisfied by this function.
Let us define by $g^<_n(r,r')$ the solution of (\ref{12}) regular at $r \to 0$,
and by $g^>_n(r,r')$ the solution that vanishes at infinity. These two solutions
must satisfy the continuity condition at $r=r'$, with their first derivative
discontinuous at this point.

It is of our main interest to investigate the vacuum polarization effect for 
external points to the magnetic flux. As we shall see, because the 
information about the field in this region must taking into account, the above 
mentioned effect will present contribution coming from the inner region. Now, 
let us consider solutions of (\ref{12}) with both $r$ and $r'$ greater than 
$R$.

Let us first consider the models ({\it{i}}) and ({\it{ii}}). The inner 
solution, $g^<_n(r,r')$, of 
(\ref{12}), corresponds to $r<r'$. However two regions must be considered separately: from $0$ to $R$ and from $R$ to $r^{'}$. Integrating out in region 
$r<R$, we have:
\begin{eqnarray}
g^<_n(r,r')=\left\{\begin{array}{cc}
A_n^{i}H_{i}(r),&\mbox{for $r<R$} \\
B^i_{n}I_{|\nu|}(\beta r)+C^i_{n}K_{|\nu|}(\beta r),
&\mbox{for $R<r<r^{'}$,}
\end{array}
\right.
\label{14}
\end{eqnarray}
where 
\begin{equation}
\nu=\frac{(n-\delta)}{\alpha} \ ,
\end{equation} 
being 
\begin{equation}
\delta=\frac{e\phi}{2\pi \alpha}=N+\gamma \ .
\end{equation} 
$H_{i}(r)$, for $i=1,2$, represents the solution associated with the two firsts 
models:
\begin{equation}
H_{1}(r)=\frac{1}{r}M_{\sigma_{1}, \lambda_{1}}\left( \frac{\delta}{\alpha R^2} 
r^2\right),
\label{15}
\end{equation}
with $\sigma_{1}=(\frac{n}{\alpha}-\frac{\beta^2 R^2\alpha}{2\delta})/2$ and 
$\lambda_{1}=n/2\alpha$, and 
\begin{equation}
H_{2}(r)=\frac{1}{\sqrt{r}}M_{\sigma_{2}, \lambda_{2}}\left( \zeta r\right),
\label{16}
\end{equation}
with $\sigma_{2}=\frac{n\delta}{\alpha}(\delta^2-\beta^2 R^2 \alpha^2)^{-1/2}$, 
$\lambda_{2}=n/\alpha$ and $\zeta=\frac{2}{R\alpha}(\delta^2+\beta^2 R^2 
\alpha^2)^{1/2}$. 

In both definitions above, Eq.s (\ref{15}) and (\ref{16}), 
$M_{\sigma, \lambda}$, represents the Whittaker function \cite{tabela}. The continuity condition of $g_{n}^{<}$ and its first derivative at $r=R$, allow us to obtain the ratio $D^i_n=C^i_{n}/B^i_{n}$ as a function of $\beta R$ only:
\begin{equation}
D_n^i(\beta R)=\frac{H_i'(R)I_{|\nu|}(\beta R)-H_i(R)I'_{|\nu|}(\beta R)}
{H_{i}(R)K^{'}_{|\nu|}(\beta R)-H_i'(R)K_{|\nu|}(\beta R)} \ .
\label{17}
\end{equation} 
In all the expressions, $I_{|\nu|}$ and $K_{|\nu|}$ are the modified Bessel 
functions.

The outer solution of (\ref{12}) is given by
\begin{equation}
g^{>}_{n}(r,r^{'})=E_n^{i}K_{|\nu|}(\beta r), 
\quad \hbox{for $r>r^{'}$.}
\label{18} 
\end{equation}

All the above constants are determined by imposing the boundary conditions
explained before on $g_n^<$ and $g_n^>$ at $r=r'$. In this way they will be
function of $r'$. Finally we obtain the result to the complete radial
Green function below:
\begin{eqnarray}
g_n(r,r^{'})=-\frac{1}{\alpha}K_{|\nu|}(\beta r_{>})\left( I_{|\nu|}(\beta r_{<})+D^i_n K_{|\nu|}(\beta r_{<}) \right) \quad \hbox{for $r$ and $r^{'}>R$.}
\end{eqnarray}
In the above equation $r_>$ $(r_<)$ is the bigger (smaller) value between
$r$ and $r'$. Substituting this solution into (\ref{11}) and after some 
intermediate steps, the Euclidean Green function acquires the following 
expression:
\begin{eqnarray}
G(x,x')&=&\frac{e^{i N\Delta\theta}}{8\pi^2\alpha rr'\sinh u_0}
\frac{e^{i\Delta \theta}\sinh(\gamma u_0/\alpha)+\sinh[(1-\gamma)u_0/
\alpha]}{\cosh(u_0/\alpha)-\cos\Delta\theta}
\nonumber\\
&+&\frac{1}{4 \pi^{2}}\int_0^{\infty} d\beta \beta J_{0}\left(\beta\sqrt{(\Delta 
\tau)^{2}+(\Delta z)^{2}}\right)\times
\nonumber\\
&&\sum_{n=-\infty}^{\infty}e^{i n\Delta\theta}D^i_n(\beta R)K_{|\nu|}(\beta r)K_{|\nu|}(\beta r') ,
\label{19}
\end{eqnarray}
where
\begin{equation}
\cosh u_{o}=\frac{r^2+{r^{'}}^{2}+(\Delta \tau)^{2}+(\Delta z)^{2}}{2rr^{'}} \ .
\label{20}
\end{equation}

We can observe that the first term of the above Green function is, up to a gauge 
transformation, equivalent to the result presented by Guimar\~aes and 
Linet \cite{linet1} for the 
massless scalar field. The second term presents, besides the dependence on the 
$\alpha$ parameter and the magnetic flux, a dependence on the radius $R$ of 
the tube through the constant $D^{i}_n$. Moreover, this Green function obeys 
the boundary condition $G(\theta'=\theta+2\pi)=G(\theta'=\theta)$, which is a
consequence of having used explicitly the vector potential in the differential
equation (\ref{Gr}). Although we have calculated this Green function for points
outside the magnetic tube, there remains contributions coming from this region
that do not disappear by taking $\gamma=0$ and $\Delta\theta=2\pi$.

As to the third model, the solution to $g_{n}(r,r')$ is:
\begin{eqnarray}
g^<_n(r,r')=\left\{\begin{array}{cc}
A_n I_{|\nu|}(\beta r),&\mbox{for $r<R$}\\
B_{n}I_{|n/\alpha|}(\beta r)+C_{n}K_{|n/\alpha|}(\beta r),
&\mbox{for $R<r<r^{'}$,}
\end{array}
\right.
\label{21}
\end{eqnarray}
and
\begin{equation}
g^>_n(r,r')=E_n K_{|n/\alpha|}(\beta r), 
\quad \hbox{for $r>r^{'}$.}
\label{22} 
\end{equation}

Once more, the complete solution to this radial function can be obtained adopting a similar procedure as the previous case. This solution is:

\begin{eqnarray}
g_n(r,r^{'})=-\frac{1}{\alpha}K_{|n|/\alpha}(\beta r_{>})\left( I_{|n|/\alpha}(\beta r_{<})+D_n K_{|n|/\alpha}(\beta r_{<}) \right) \nonumber \\ 
\quad \hbox{for $r$ and $r^{'}>R$.} 
\end{eqnarray}
Again imposing the continuity condition on (\ref{21}) and its first derivative at 
$r=R$ we get $D_n=C_n/B_n$ as function of $\beta R$:
\begin{equation}
D_{n}(\beta R)=\frac{I_{|\nu|}^{'}(\beta R)I_{|n|/\alpha}(\beta R)-I_{|\nu|}(\beta R)
I^{'}_{|n|/\alpha}(\beta R)}{I_{|\nu|}(\beta R)K^{'}_{|n|/\alpha}(\beta R)-
I_{|\nu|}^{'}(R)K_{|n|/\alpha}(\beta R)} \ .
\label{23}
\end{equation}

Finally adopting similar procedure as in the previous case, we obtain:
\begin{eqnarray}
G(x,x')&=&\frac1{8\pi^2 \alpha rr'\sinh u_0}\left[\frac{\sinh(u_0/\alpha)}
{\cosh(u_{o}/\alpha)-\cos\Delta\theta}\right]
\nonumber\\
&+&\frac1{4 \pi^{2}}\int_0^{\infty} d\beta \beta J_{0}\left(\beta\sqrt{(\Delta 
\tau)^{2}+(\Delta z)^{2}}\right)\times
\nonumber\\
&&\sum_{n=-\infty}^\infty e^{in\Delta\theta}
D_n(\beta R)K_{|n|/\alpha}(\beta r)K_{|n|/\alpha}(\beta r') \ .
\label{24}
\end{eqnarray}

For this case we can see that the first term in this Green function coincides 
with results found by Smith \cite{smith} and by Linet in \cite{linet2} for a 
massless scalar field. Moreover, the second term contains the information 
about the radius of the tube through the constant $D_n$.

So having the Green functions our next steps will be the calculations of the 
vacuum expectation values associated with the modulus of the scalar field and 
its energy-momentum tensor.

\section{Computation of $\langle\hat{\Phi}^*(x)\hat{\Phi}(x)\rangle_{Ren.}$}

In this section we want to calculate the renormalized vacuum 
expectation values associated with the modulus square of the scalar field in this 
physical system for the three distinct configurations of magnetic fields. As we
shall see these vacuum average will depend on the conicity parameter $\alpha$, 
the fractional part of $\phi/\phi_0$ and the radius of the tube, $R$. Because 
the two firsts models present similar Green functions, we shall develop
initially  the calculations considering these two models together and later the third one.

The renormalized vacuum expectation value of the modulus square of the charged massless 
scalar field on idealized cosmic string spacetime is given by
\begin{equation}
\langle\hat{\Phi}^*(x)\hat{\Phi}(x)\rangle_{Ren.}=\lim_{x'\to x}\left[G(x,x')-
G_{E}(x,x')\right],
\label{25}
\end{equation}
where $G_{E}(x,x')=\frac1{4\pi^2(x-x')^{2}}$, is the Euclidean Green function for 
a flat four-dimensional spacetime, and $G(x,x')$ the Euclidean Green function 
associated with this system under investigation.

As we have explained before, we shall develop first this calculation for the 
models ({\it{i}}) and ({\it{ii}}), whose Green functions are given in (\ref{19}) 
with $D^i_n$ given by (\ref{17}). A very interesting aspect of these Green 
functions is that the correction due to the magnetic tube's radius is finite in the coincidence limit; so the result is:
\begin{eqnarray}
\langle\hat{\Phi}^*(x)\hat{\Phi}(x)\rangle_{Ren.}&=&\frac1{48\pi^2r^ 2\alpha^2}
[1-\alpha^2 -6\gamma(1-\gamma)]
\nonumber\\
&+&\frac1{(2\pi)^{2}}\int_0^{\infty} d\beta \beta \sum_{n=-\infty}^\infty
D^i_n(\beta R)K^{2}_{|\nu|}(\beta r).
\label{26}
\end{eqnarray}  
In order to provide a quantitative information about the behavior of the correction in above equation, we must adopt the following procedure: ${\it{i )}}$ Redefining a new variable $v=\beta r$, it is possible to see that the second term presents on overall $1/r^2$ dependence, the same as presented by first term. However there remains an extra radial dependence on the integrand through the coefficient $D_n^i(vR/r)$. So the complete information about the second term can only be obtained numerically specifying the ratio $R/r$. ${\it{ii )}}$ An approximated behavior can be exhibited if we analyse the integrand of (\ref{26}). Near the origin $K_{|\nu|}(v)$ behave as $v^{-2|\nu|}$, on the other hand $D_n^{i}(vR/r)$ present a dependence $v^{2|\nu|}$ as we shall see later. So the product $vD_n^{i}(vR/r)K^{2}_{|\nu|}(v)$ vanishes linearly with $v$. For large $v$, $vK^{2}_{|\nu|}(v)$ decays as $e^{-2v}$. Because $D_n^{i}(vR/r)$ grows slowly than $e^{2vR/r}$ for large $v$, its most significant relevance in the integrand 
happens only up $v$ of order unity. However because we are mostly interested to study the vacuum polarization for points very far from the cosmic string, we shall consider $R/r\ll 1 $ in the our analysis. So we may approximate the integrand of (\ref{26}) assuming to the coefficient $D_n^i$ its first order expansion. This is
equivalent to consider the coefficients $D^i_n(\beta R)$ with $\beta R \ll 1$. 
Taking in (\ref{17}) $\beta R\ll 1$, the most relevant term for both cases are
given by
\begin{equation}
D^i_n(\beta R)=-\frac{2}{\Gamma(|\nu|+1)\Gamma(|\nu|)}\left(\frac{g^n_i
-|\nu|f^{n}_{i}}{g^{n}_{i}+|\nu|f^{n}_{i}}\right)\left( \frac{\beta R}{2}
\right)^{2|\nu|},
\label{27}
\end{equation} 
where
\begin{equation}
g^{n}_{1}=\left( \frac{\delta-n}{\alpha}-1\right)M_{\lambda_1,\lambda_2}
(\delta/\alpha)+\left( 1+\frac{2n}{\alpha}\right)M_{\gamma_1,\lambda_1}
(\delta/\alpha),
\label{28}
\end{equation}
\begin{equation}
f^{n}_{1}=M_{\lambda_{1},\lambda_{2}}(\delta/\alpha),
\end{equation}
\label{29}
\begin{equation}
g^{n}_{2}=\left( \frac{\delta-n}{\alpha}-\frac{1}{2}\right)M_{\lambda_2,
\lambda_1}(2\delta/\alpha)+\left(\frac12+\frac{2n}{\alpha}\right)M_{\gamma_2,
\lambda_2}(2\delta/\alpha) \ ,
\label{30}
\end{equation}
and
\begin{equation}
f^n_2=M_{\lambda_2,\lambda_1}(2\delta/\alpha) \ ,
\label{31}
\end{equation}
being $\gamma_1=(n+2\alpha)/2\alpha$, $\gamma_2=(n+\alpha)/\alpha$ and
$\nu=\frac{n-N-\gamma}\alpha.$

Using now these approximated expressions to the coefficients $D^i_n$, and with the 
help of \cite{tabela}, Eq. (\ref{26}) becomes:
\begin{eqnarray}
\langle\hat{\Phi}^*(x)\hat{\Phi}(x)\rangle_{Ren.}&=&\frac1{48\pi^2r^2\alpha^2}
[1-\alpha^{2}-6\gamma(1-\gamma)]
\nonumber \\
&-&\frac1{(2\pi)^2\alpha r^2}\sum_{n=-\infty}^\infty\frac{|\nu|}{1+2|\nu|}
\left(\frac{g^n_i-|\nu|f^n_i}{g^n_i+|\nu|f^n_i}
\right)\left(\frac Rr\right)^{2|\nu|} \ .
\label{32}
\end{eqnarray}

The first term of the above result depends only on the fractional part of
$\phi/\phi_0$, besides the parameter $\alpha$. On the other hand the second term,
the correction, depends on the integer part of the the ratio $\phi/\phi_0$, 
$N$. However, because $R/r$ is assumed to be smaller than unity, the most 
relevant correction comes from the $n=N$. In this case the extra radial dependence
of the second term is dominated by $(R/r)^{2\gamma/\alpha}$.

As we have already explained our approximation consists to substitute into (\ref{26}), the expression 
(\ref{27}) to the coefficients $D^{i}_{n}$. In fact doing this other contributions 
given by the complete expression (\ref{23}) are missing; only the first order in the 
expansion was kept. In order to be sure that our approximation is a good one, 
we also analysed numerically the integrand of the second contribution of (\ref{26}) 
using the exact expression (\ref{23}) and the approximated result 
(\ref{27})-(\ref{31}) for the coefficients for both cases $i=1$ and $2$. For simplicity only
we take in the summation $N=0$, for both cases. This approximation is once more justified by the figures $1(a)$ and $1(b)$ which exhibit comparatively the integrand of (\ref{26}) written in terms os the new variable $v$, $I(v)=v\sum D_n^i(vR/r)K^2_{|\nu|}(v)$, for the models $(i)$ and $(ii)$, respectively, considering the complete expressions for the coefficients and their respective approximations. We can see a good agreement between both curves for each case. (In 
these numerical analysis we have considered only the term $n=0$ in the 
summation for both models. Considering other terms, for example $n$ running 
from $-20$ to $20$, we have not observed any significant difference when
compared with the $n=0$ term. So the summation in (\ref{32}) can be mainly 
represented by its most relevant contribution.)

So accepting this fact as general, our final result to the vacuum polarization 
effects becomes:
\begin{eqnarray}
\langle\hat{\Phi}^*(x)\hat{\Phi}(x)\rangle_{Ren.}&=&\frac1{48\pi^2r^2\alpha^2}
\left\{\frac{1-\alpha^2-6\gamma(1-\gamma)}{\alpha^2}\right.
\nonumber \\
&-&\left.\frac{12\gamma}{\alpha(2\gamma+\alpha)}
\left[\frac{g^N_i-\gamma/\alpha f^N_i}{g^N_i+\gamma/\alpha f^N_i}\right]
\left(\frac{R}{r} \right)^{2\gamma/\alpha}\right\} \ .
\label{33}
\end{eqnarray}
From the above expression, it is possible see that the second term has the same 
order of magnitude as the standard one, up to ratio $r/R=
|\overline{x}|^{\alpha/2\gamma}$, being $\overline{x}$ given by
\begin{equation}
\overline{x}=\left[\frac\alpha{1-\alpha^2+6\gamma (\gamma-1)}\frac{12\gamma}
{\alpha+2\gamma}\left(\frac{g^N_i-\gamma/\alpha f^N_i}{g^N_i+\gamma/\alpha 
f^N_i}\right)\right] \ .
\end{equation}
For smaller ratio the second contribution becomes more relevant than 
the first one. Once more, by numerical analysis we observe that for $N=0$,
$\alpha=0.99$ and $\gamma=0.3$, there happen a cancellation between both
contributions for $r/R$ of order $5$ for both models. For $N=1$ and 
assuming the same values for $\alpha$ and $\gamma$, the seconds terms changes 
their signs and there is no cancellation anymore. 

After this discussion about the two firsts models, let us analyze the third one 
where the respective Green function is given by (\ref{24}). For this case the 
renormalized vacuum expectation value of the modulus square of the scalar field 
becomes:
\begin{eqnarray}
\langle\hat{\Phi}^*(x)\hat{\Phi}(x)\rangle_{Ren.}&=&\frac1{48\pi^2r^2\alpha^2}
(1-\alpha^2)
\nonumber \\
&+&\frac1{(2\pi)^2}\int_0^{\infty} d\beta \beta \sum_{n=-\infty}^\infty 
D_n(\beta R)K^2_{|n|/\alpha}(\beta r) \ ,
\label{35}
\end{eqnarray}
with the coefficients $D_{n}(\beta R)$ given by (\ref{23}). Although the 
physical system that we are analysing is 
completely different from the one analyzed by Bruce Allen {\it {et al}} in 
\cite{allen}, there appears a strong similarity between our result with theirs.  
Again if we are inclined to adopt an approximate expression to the coefficients 
$D_n$ considering $\beta R \ll 0$, two distinct behavior are obtained:
\noindent\\
For $n \not= 0$, 
\begin{equation}
D_n(\beta R)=-\frac2{\Gamma(|n|/\alpha+1)\Gamma(|n|/\alpha)}
\left(\frac{|\nu|-|n|/\alpha}{|\nu|+|n|/\alpha}\right)
\left(\frac{\beta R}{2}\right)^{2|n|/\alpha},
\label{36}
\end{equation} 
and for $n = 0$, and
\begin{equation}
D_0(\beta R)=\frac{\delta/\alpha}{\delta/\alpha\left[\ln\left(\frac{\beta R}2
\right)+{\mathcal{C}}\right]-1} \ ,
\label{37}
\end{equation}
where ${\mathcal{C}}$ is the Euler constant. As we can see for $n \not= 0$, 
$D_n(\beta R)$ vanishes at least as fast as $(\beta R)^{2|n|/\alpha}$ when 
$\beta R \rightarrow 0$. On the other hand, $D_{0}(\beta R)$ vanishes only with 
the inverse of the logarithm, so slowly. In this way the most relevant 
contribution to the summation in (\ref{35}) comes from the $n=0$. This same 
conclusion was previously obtained by Bruce Allen {\it {et al}} for another 
physical system. There they used a more convenient notation to express this 
coefficient, which also we adopt here: 
\begin{equation}
D_0(\beta R)=\frac1{\ln(\beta/q)},
\label{38}
\end{equation}
where in our case
\begin{equation}
q=\frac2{R}e^{-{\mathcal{C}}+\alpha/\delta}.
\label{39}
\end{equation}

So the most relevant contribution to the correction of the renormalized vacuum 
expectation value of the square of modulus of the scalar field in (\ref{35}) is 
proportional to:
\begin{equation}
\int_0^{\infty} d\beta \beta \frac{K^2_0(\beta r)}{\ln(\beta/q)}=\frac1{r^2}
\int_0^{\infty} dv v \frac{K^2_0(v)}{\ln(v)-\ln(qr)} \ .
\label{40}
\end{equation}

As it was pointed out in \cite{allen} this approximation presents a problem because
the above integrand has a pole at $v=qr$. However, this pole is a consequence the approximation adopted, and it occurs when the argument of $D_0$ is much 
greater than unity, where the approximation is no longer valid. Moreover analyzing the 
behavior of the integrand, they observe that $vK^{2}_{0}(v)$ vanishes at $v=0$, 
peaks around $v$ of order $1$, and decays as $e^{-2v}$ for large $v$. On the other 
hand $D_0$ vanishes for $v=0$ and grows less slowly than $e^{2vR/r}$ for large $v$. So, 
finally they conclude that (\ref{40}) can be well approximated discarding the 
term $\ln(v)$. (Also a numerical analysis in their paper supports this 
approximation.)

Here in our case a similar conclusion can be reached. In fact choosing 
$\gamma=0.2$, 
$N=0$ and $\alpha=0.99$, $qr=\frac{2r}Re^{-{\mathcal{C}}+\alpha/\delta}$ in of 
order than $10^{5}$ for $r/R=10^{3}$, so $\beta R \approx 10^2$. Consequently the pole occurs in the region where the approximation is no longer valid. We shall not repeat their argument to accept the same approximation procedure; however we present in Fig. $2$ a numerical evaluation of this approximation, 
comparing the integrand of the second contribution of (\ref{35}) using the
exact expression to its more relevant term, $D_0$, given by (\ref{23}) and its
approximated expression given in (\ref{40}), discarding the $\ln(v)$ term. So 
we can conclude, supported by those evidences, that the second term of 
(\ref{35}) can be approximated by
\begin{equation}
-\frac1{(2\pi)^2}\frac1{r^2\ln(qr)}\int_0^{\infty} dv v K^2_0(v).
\label{41}
\end{equation}
Finally with the help of \cite{tabela} we integrate the above equation, providing 
to (\ref{35}) the expression below:
\begin{equation}
\langle\hat{\Phi}^*(x)\hat{\Phi}(x)\rangle_{Ren.}=\frac1{48\pi^2r^2}\frac{1-\alpha^2
}{\alpha^2}
-\frac1{8\pi^2\alpha r^2\ln(\frac{2r}R e^{-{\mathcal{C}}+\alpha/\delta})} \ .
\label{42}
\end{equation}

From the above expression we can see that the explicit dependence on the flux tube 
appears only on the second term. Taking $R \rightarrow 0$ or $\delta \rightarrow 0$,
this term vanishes. (The first term contains only the contribution coming from the 
geometry and vanishes when $\alpha \rightarrow 1$).

As in the physical system analyzed in \cite{allen}, there is in this case a long-range effect 
in (\ref{42}) coming from the special structure of the magnetic tube around the 
cosmic string.

Because $\alpha \approx 1$, (\ref{42}) can be written as
\begin{eqnarray}
\langle\hat{\Phi}^*(x)\hat{\Phi}(x)\rangle_{Ren.}=\frac{(\alpha^{-1}-1)}
{24\pi^2r^2}\left[1-\frac{3}{(\alpha^{-1}-1)\ln(\frac{2r}Re^{-{\mathcal{C}}+
\alpha/\delta})}\right] \ ,
\label{43}
\end{eqnarray}
where we see that the second term inside the bracket vanishes slowly as $r \rightarrow \infty$. In fact it can be noticed that the second term is greater than the unity up to
\begin{equation}
\frac rR \approx\frac12\exp\left[\frac3{\alpha^{-1}-1}+{\mathcal{C}}-\frac{\alpha}
{\delta}\right],
\label{44}
\end{equation}
which $\alpha=0.99$ and $\delta=0.2$ we have $r/R=10^{127}$. As in \cite{allen}, even for $R=10^{-30} cm$, the correction in (\ref{43}) becomes more relevant than 
the geometrical effect up to $r \approx 10^{97} cm$ which exceed the radius of the observable Universe.

\section{Computation of $\langle\hat{T}_\mu^\nu(x)\rangle_{Ren.}$}

The renormalized vacuum expectation value of the energy-momentum 
tensor associated with this system under investigation is given by:
\begin{equation}
\langle\hat{T}_\mu^\nu\rangle_{Ren.}=\lim_{x'\to x}[D_{\mu(\alpha,\phi)}^{\nu'}
G(x,x')-D_{\mu(1,0)}^{\nu'}G_{E}(x,x')] \ ,
\label{45}
\end{equation}
with the operator $D_{\mu(\alpha,\phi)}^{\nu'}$ expressed as
\begin{equation}
D_{\mu(\alpha,\phi)}^{\nu'}=(1-2\xi)D_\mu{\overline{D}}^{\nu'}-\xi (D_\mu D^\nu
+\overline{D}_{\mu'}\overline{D}^{\nu'})+(2\xi-\frac12)\delta_\mu^\nu D_{\sigma}
\overline{D}^{\sigma'} \ ,
\label{46}
\end{equation}
being $D_\sigma=\nabla_\sigma-ieA_\sigma$ and the bar denoting complex 
conjugate. Again, in order to take into account the presence of the three 
magnetic field configurations given in the beginning of this paper we write the four vector potential in the form $A_\sigma=(0,0,\frac{\phi a(r)}{2\pi\alpha},0)$, 
with $a(r)$ being given by (\ref{9}) and (\ref{10}), for the firsts two cases and 
by (\ref{10.1}) for the third one. The respective Green functions 
$G(x,x')$, are given by (\ref{19}) and (\ref{24}).

Because we are convince ourselves about the approximation procedure adopted in the
last section, here we shall consider in all the calculations only the leading term 
of the coefficients $D^i_n$ for each 
case.

Let us start with the two firsts models. After some intermediate calculations, we
arrive at:
\begin{eqnarray}
\langle\hat{T}_0^0(x)\rangle_{Ren.}&=&\langle\hat{T}_0^0(x)\rangle_{Reg.}+\frac1
{4\pi^2\alpha}\int_0^\infty d\beta\beta D^i_{n=N}(\beta R)\left\{2\beta^{2}
\left(\xi-\frac14\right)\times\right.
\nonumber\\
&&\left.K_{\gamma/\alpha+1}^{2}(\beta r)+
\left(2\xi \beta^2+4\left(\xi-\frac14\right)\right)\frac{\gamma^2}
{\alpha^2r^2}K_{\gamma/\alpha}^2(\beta r)\right.
\nonumber\\
&-&\left.4\left(\xi-\frac14\right)K_{\gamma/\alpha+1}(\beta r)K_{\gamma/\alpha}
(\beta r)\right\} \ , 
\label{47}
\end{eqnarray}
\begin{eqnarray}
\langle\hat{T}_1^1(x)\rangle_{Ren.}&=&\langle\hat{T}_1^1(x)\rangle_{Reg.}+\frac1
{4\pi^2\alpha}\int_0^{\infty} d\beta \beta D^i_{n=N}(\beta R)
\left\{\frac{\beta^2}{2}K^2_{\gamma/\alpha+1}(\beta r)\right. 
\nonumber \\
&-&\left(2\xi+\frac{\gamma}{\alpha}\right)\frac{\beta}{r}
K_{\gamma/\alpha+1}(\beta r)K_{\gamma/\alpha}(\beta r)
\nonumber\\
&-&\left.\left(\frac{\beta^2}2-\frac{2\xi\gamma}{\alpha r^2}\right)
K^2_{\gamma/\alpha}(\beta r) \right\} \ ,
\label{48}
\end{eqnarray}
\begin{eqnarray}
\langle\hat{T}_2^2(x)\rangle_{Ren.}&=&\langle\hat{T}_2^2(x)\rangle_{Reg.}+\frac1
{4\pi^{2}\alpha}\int_0^{\infty} d\beta \beta D^i_{n=N}(\beta R)
\left\{\left(2\beta^2\xi-\frac{\beta^2}2\right)\times\right.
\nonumber\\
&&\left.K^2_{\gamma/\alpha+1}(\beta r)+\left[2\xi-\frac{4\gamma}{\alpha}
\left(\xi-\frac{1}{4}\right)\right]\frac{\beta}{r}K_{\gamma/\alpha+1}
(\beta r)K_{\gamma/\alpha}(\beta r)\right.
\nonumber\\
&+&\left.\left[ 4\beta^2\left(\xi-\frac14\right)+
\frac{4\xi\gamma}{\alpha r^2}\left(\frac{\gamma}{\alpha}-\frac12\right)\right]
K^2_{\gamma/\alpha}(\beta r)\right\} \ ,  
\label{49}
\end{eqnarray}
and by boost invariance in the $t-z$ plane $\langle\hat{T}_3^3(x)\rangle_{Ren.}
=\langle\hat{T}_0^0(x)\rangle_{Ren.}$. Where the expression to 
$\langle\hat{T}_\mu^\nu(x)\rangle_{Reg.}$ is given by
\begin{equation}
\langle\hat{T}_\mu^\nu(x)\rangle_{Reg.}=u_1(\alpha,\gamma)diag(1,1,-3,1)
+\left(\xi-1/6\right)u_2(\alpha,\gamma)diag(1,-1/2,3/2,1) \ ,
\label{50}
\end{equation}
with
\begin{equation}
u_1(\alpha, \gamma)=\frac1{1440\pi^2r^4}[(\alpha^{-4}-1)-30\alpha^{-4}
\gamma(1-\gamma^2)]
\label{51}
\end{equation}
and
\begin{equation}
u_2(\alpha, \gamma)=\frac1{2\pi^{2}r^{4}}\left[\frac16(\alpha^{-2}-1)+
\alpha^{-2}\gamma(\gamma-1)\right].
\label{52}
\end{equation}
Substituting the asymptotic expression for $D^i_{n=N}(\beta R)$, Eq.s (\ref{27})-
(\ref{31}), into the above equations, after long calculation, we obtain:
\begin{eqnarray}
\langle\hat{T}_\mu^\nu(x)\rangle_{Ren.}&=&\langle\hat{T}_\mu^\nu(x)\rangle_{Reg.}
+\frac{\gamma}{4\pi^2r^4}\left(\frac Rr\right)^{2\gamma/\alpha}
\left(\frac{g_i^N-\gamma/\alpha f_i^N}{g_i^N+\gamma/\alpha f_i^N}\right)\times 
\nonumber \\
&&\frac1{\alpha^3(\alpha+2\gamma)(3\alpha+2\gamma)}\left[(\xi-1/6)
diag(v_{1},v_{2},v_{3},v_{1})\right. 
\nonumber \\
&+&\left.\frac\gamma 3\left(\gamma-\frac{\alpha}2\right)
diag(h_{1},h_{2},h_{3},h_{1})\right] \ ,
\label{53}
\end{eqnarray}
where
\begin{equation}
v_1=-(12\alpha^3+4\alpha\gamma(8\alpha+7\gamma)+8\gamma^3) \ ,
\end{equation}
\begin{equation}
v_2=6\alpha^3+2\alpha\gamma(5\alpha+2\gamma) \ ,
\end{equation}
\begin{equation}
v_3=-[6\alpha^{3}+2\alpha \gamma(5\alpha+2\gamma)](3\alpha+2\gamma) \ ,
\end{equation}
and
\begin{equation}
h_1=2\alpha+2\gamma \ ,
\end{equation}
\begin{equation}
h_2=2\alpha \ ,
\end{equation}
\begin{equation}
h_3=-6\alpha-4\gamma \ .
\end{equation}

From the above expressions we can see that: {\it{i)}} It reproduces the Guimar\~aes 
and Linet's result in the limit $R \rightarrow 0$, {\it{ii)}} the corrections 
decrease with the power of $(R/r)$ proportional to the fractional part of 
$\phi/\phi_0$ and {\it{iii)}} for $\xi=1/6$ the trace of (\ref{53}) vanishes.

Finally it should be mentioned that this expression satisfies the conservation 
condition
\begin{equation}
\nabla_\nu\langle\hat{T}_\mu^\nu(x)\rangle_{Ren.}=0 \ .
\end{equation} 

As to the third model the respective Green function is given by (\ref{24}) and 
the radial function $a(r)$ by (\ref{10.1}). Adopting a similar procedure as in 
the last two previous calculations we obtain:
\begin{eqnarray}
\langle\hat{T}_0^0(x)\rangle_{Ren.}&=&\langle\hat{T}_0^0(x)\rangle_{Reg.}
+\frac1{4\pi^{2}\alpha}\int_0^{\infty} d\beta\beta^3D_0(\beta R)
\left\{2\xi K_0^2(\beta r)\right. 
\nonumber \\
&+&\left. 2\left(\xi-\frac14\right)K_1^2(\beta r)\right\} \ , 
\label{54}
\end{eqnarray}
\begin{eqnarray}
\langle\hat{T}_1^1(x)\rangle_{Ren.}&=&\langle\hat{T}_1^1(x)\rangle_{Reg.}+
\frac1{4\pi^2\alpha}\int_0^{\infty} d\beta \beta^3D_0(\beta R)
\left\{\frac12 K_1^2(\beta r)\right. 
\nonumber \\
&-&\left.\frac{2\xi}{\beta r}K_0(\beta r)K_1(\beta r)-\frac12K_0^2(\beta r) 
\right\}, 
\label{55}
\end{eqnarray}
\begin{eqnarray}
\langle\hat{T}_2^2(x)\rangle_{Ren.}&=&\langle\hat{T}_2^2(x)\rangle_{Reg.}
+\frac1{4\pi^{2}\alpha}\int_0^{\infty} d\beta \beta^3D_0(\beta R)
\left\{2(\xi-1/4) K_1^2(\beta r) \right. 
\nonumber \\
&+&\left.\frac{2\xi}{\beta r}K_0(\beta r)K_1(\beta r)+2(\xi-1/4))K_0^2
(\beta r)\right\} \ , 
\label{56}
\end{eqnarray}
with $\langle\hat{T}_3^3(x)\rangle_{Ren.}=\langle\hat{T}_0^0(x)\rangle_{Ren.}$. 
Now we have
\begin{eqnarray}
\langle\hat{T}_\mu^\nu(x)\rangle_{Reg.}&=&\frac1{1440\pi^2r^4}(\alpha^{-4}-1)
diag(1,1,-3,1)
\nonumber \\
&+&\frac1{24\pi^2r^4}(\alpha^{-2}-1)(\xi-1/6)diag(2,-1,3,2) \ ,
\end{eqnarray}
which is the result for the renormalized vacuum expectation value of the 
energy-momentum tensor associated with a massless scalar field on a 
idealized cosmic string spacetime \cite{smith, linet2}. This result, together 
with (\ref{35}), seem very curious, because although we are considering a charged 
field in the presence of a cylindrical shell of magnetic field, differently from
the two previous cases, the electromagnetic effect appears only on 
the seconds terms for both calculations; moreover this effect is of long-range type.

Finally, substituting (\ref{38}) into (\ref{54})-(\ref{56}), taking into account 
the same considerations to overcome the integral problem on vacuum expectation 
value of the square of the scalar field, we obtain:
\begin{equation}
\langle\hat{T}_\mu^\nu(x)\rangle_{Ren.}=\langle\hat{T}_\mu^\nu(x)\rangle_{Reg.}-
\frac1{4\pi^2\alpha r^4\ln(qr)}(\xi-1/6)diag(2,-1,3,2) \ .
\label{59}
\end{equation}

It is easy to see that this result is traceless when $\xi=1/6$ as conformal 
invariance demands on this spacetime; however as notice by Bruce Allen 
{\it{et al}} for $\xi \not= 1/6$, (\ref{59}) presents a small violation of the conservation condition, which vanishes for $R \rightarrow 0$, or 
$q \rightarrow \infty$. Being $\xi\not=1/6$, the energy density associated
with a massless scalar charged field in this spacetime presents a long-range
corrections arising from the magnetic field. For realistic models where the
parameter $\alpha$ is very close to the unity this correction becomes
even more relevant than the contribution coming from the geometry itself
up to distance of the order the radius of the observable Universe.

\section{Concluding Remarks}

In this paper we have analyzed the vacuum polarization effects of a 
charged massless scalar field due to the locally flat cosmic string spacetime 
and a magnetic flux confined on a tube.

Three distinct configurations of magnetic flux have been considered. Among them, 
the cylindrical shell of $\delta$-function field presents a long-range effect on the vacuum polarization.

In general, the results obtained to $\langle\hat{T}_\mu^\nu(x)\rangle_{Ren.}$ 
as to $\langle\hat{\Phi}^*(x)\hat{\Phi}(x)\rangle_{Ren.}$, outside the tube, 
present corrections due to the magnetic tube which vanish when we take 
$\phi \rightarrow 0$ or $R \rightarrow 0$. 

Specifically speaking about the trace anomaly, it is well known that 
\cite {birrel} 
\begin{equation}
\langle\hat{T}_\mu^\mu(x)\rangle_{Ren.}=\frac{a_2(x)}{16\pi^2} \ ,
\label{47}
\end{equation}
where $a_2(x)$ presents contributions coming from the geometry of the 
spacetime as well as contribution from the gauge field strength \cite{jack}. 
However because the spacetime under consideration is locally flat, and the 
region of our interest has no magnetic field we have $a_{2}(x)=0$.

We conclude this paper by mentioning the importance to consider a finite structure of the magnetic flux on the quantum aspects of a charged field on an Abelian
cosmic string spacetime. Although the complete knowledge about this structure could not be very well understood, its influence on the vacuum polarization effect can be even more relevant than the influence of geometry itself.

\newpage

{\bf{Acknowledgments}}
\\       \\
We would like to thank to V. M. Mostepanenko for a critical and enthusiastic
reading of this paper. Also we would like to thank Conselho Nacional de 
Desenvolvimento Cient\'\i fico e Tecnol\'ogico (CNPq.) and CAPES for partial 
financial support.

\newpage

\section{Figure Captions}

{\bf Figure 1}: The solid curve in Fig. $1(a)$ represents the integrand of Eq. 
(\ref{26}) written in terms of the new variable $v=\beta r$ for the first 
model considering the complete expression for the coefficient $D^1_0$ given 
by (\ref{17}). The dashed curve is the same integrand now using the 
approximated expression for $D^1_0$, Eq. (\ref{27}). A similar analysis for
the second model is given in Fig. $1(b)$. In both numerical analysis we have used
$\alpha=0.99$, $\gamma=0.2$, $N=0$ and $R/r=10^{-3}$.

{\bf Figure 2}: The solid curve in Fig. $2$ represents the integrand of 
(\ref{35}) for the third model considering the exact expression for $D_0$, 
given by (\ref{23}). The dashed curve exhibits the same integrand
considering now the approximated expression to $D_0$. In this numerical
analysis we have used $\alpha=0.99$, $\gamma=0.2$, $N=0$ and $R/r=10^{-3}$. 
\newpage
\begin{figure}[t]
\begin{center}
\includegraphics[width=8cm,angle=-90]{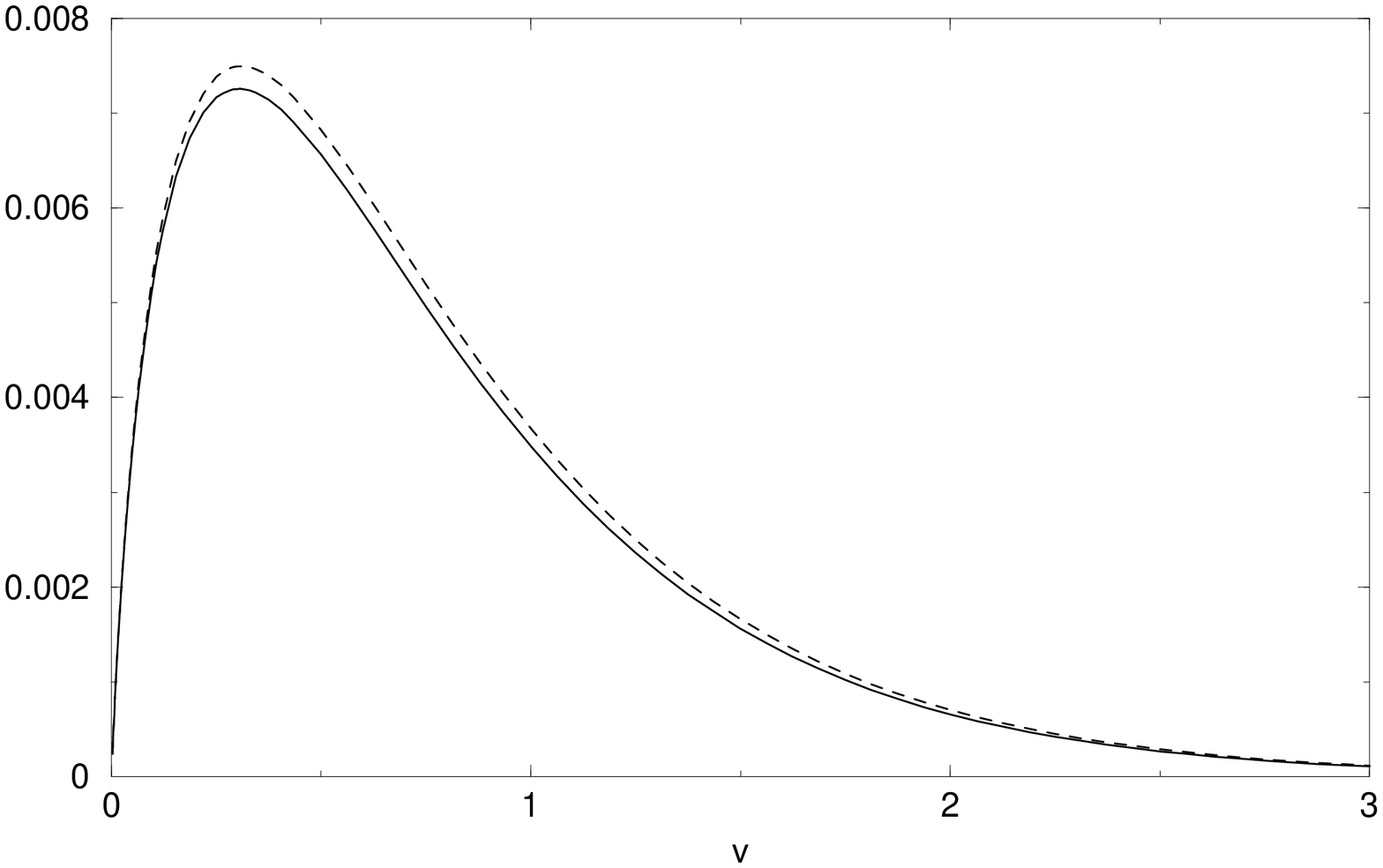}

(a)

\includegraphics[width=8cm,angle=-90]{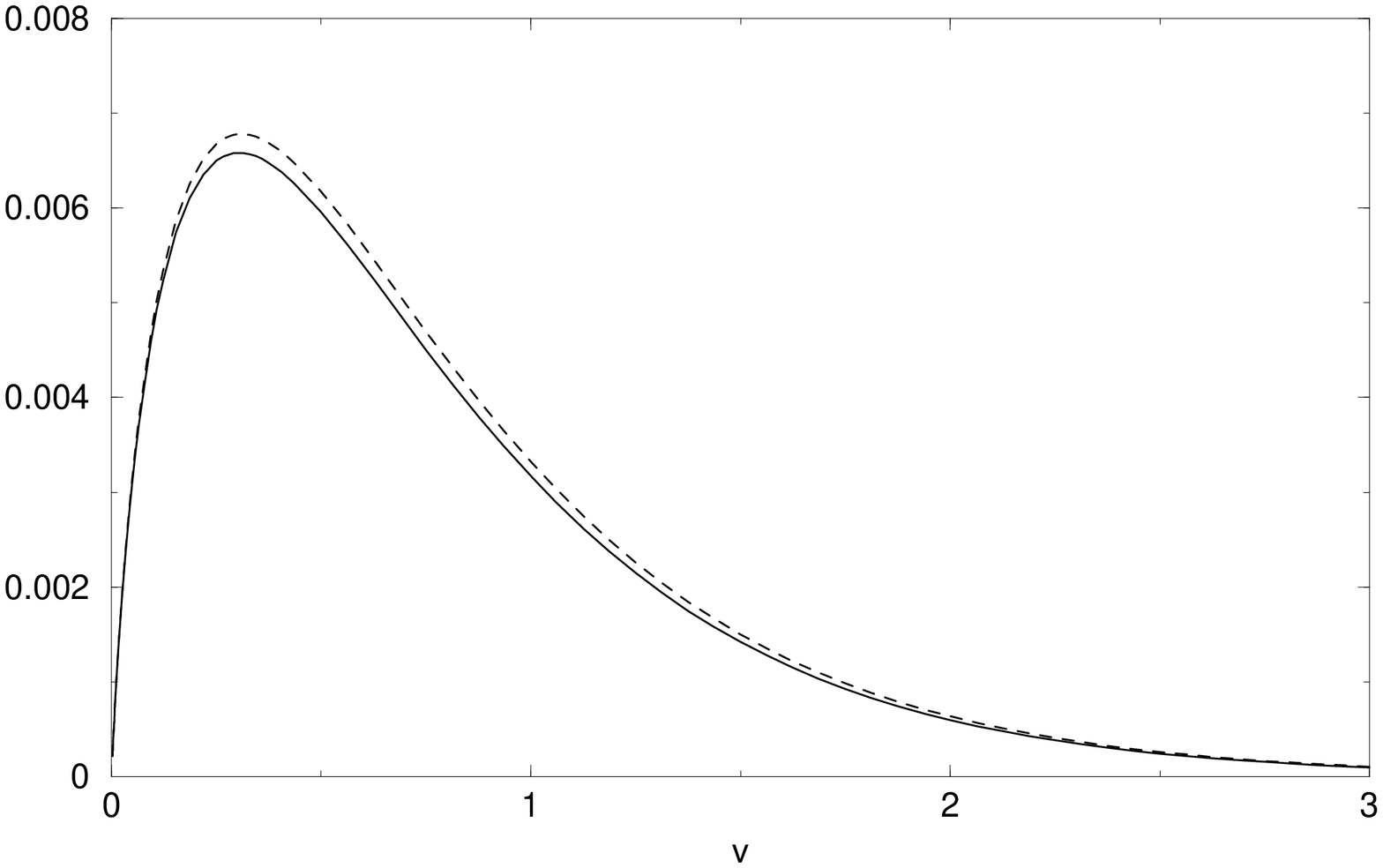}

(b)

\label{fig1}
\caption{}
\end{center}
\end{figure}

\newpage
\begin{figure}[t]
\begin{center}
\includegraphics[width=8cm,angle=-90]{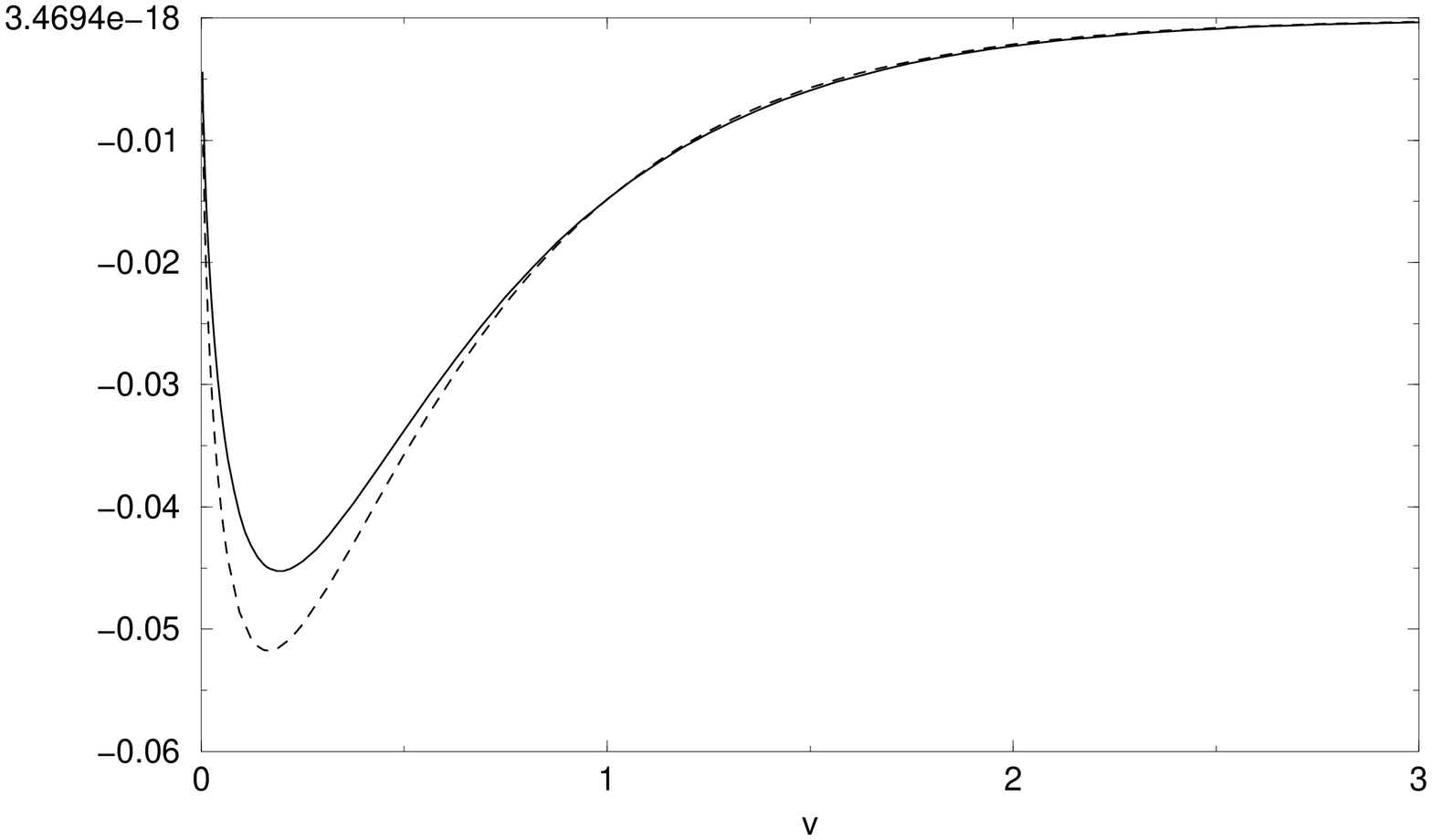}

\label{fig2}
\caption{}
\end{center}
\end{figure}

\end{document}